\documentclass{cg-sge}

\title{
\fontsize{24pt}{24pt}\selectfont
Science ouverte et collaborative pour l'élaboration d'un banc automatisé de caractérisation de pertes en commutation par opposition
}

\author{
\fontsize{11pt}{11pt}\selectfont
Nicolas ROUGER$^1$, Luiz VILLA$^{2,3}$, Matthieu MASSON$^{1,4}$, Pauline KERGUS$^1$, Joseph KEMDENG$^1$,\\
Lorenzo LEIJNEN$^{1,4}$, Jean ALINEI$^3$, Adrien COLOMB$^1$, Ayoub FARAH-HASSAN $^{2,3}$, Arnauld BIGANZOLI$^1$\\
\fontsize{10pt}{10pt}\selectfont
$^1$Laplace, CNRS, INPT, Université de Toulouse, Toulouse, France.\\
\fontsize{10pt}{10pt}\selectfont$^2$LAAS-CNRS, Toulouse, France. \\
\fontsize{10pt}{10pt}\selectfont$^3$Owntech fondation.\\
\fontsize{10pt}{10pt}\selectfont$^4$NXP Semiconductors, Toulouse, France. \\
}

\date{}

\begin{document}

\maketitle
\thispagestyle{fancy}

\fontsize{9pt}{9pt}\selectfont
\textbf{RESUME -- La mesure des pertes en commutation des transistors de puissance s'effectue généralement par la mesure dite de double impulsion. La mesure par opposition de deux cellules de commutation constitue une méthode complémentaire, plus précise mais indirecte. Néanmoins, la mise en oeuvre de cette méthode peut être plus complexe, et nécessite des étapes de calibration et un pilotage complet, avec la problématique de la gestion thermique. Dans ce contexte, nous avons proposé d'aborder ce sujet par une action de science ouverte et collaborative, d'abord sous la forme d'un hackathon de 2 jours, puis des séances mensuelles ouvertes. Plus de 20 participant.e.s ont contribué sur les 2 jours du hackathon, suivi de séances mensuelles pour les personnes souhaitant poursuivre ensemble. Cela a permis de mettre en place un banc automatisé, en science ouverte, incluant la génération des ordres de commutation, la configuration et le pilotage des instruments de mesure, et la partie hardware. Nous présentons et partageons ici nos travaux et cette approche ouverte.}\\

\textbf{\textit{Mots-clés -- Caractérisation, pertes, commutation, instrumentation, automation, science ouverte.}}

\fontsize{10pt}{10pt}\selectfont

\section{Introduction}
Le verrou scientifique qui a été l'objet de cette science ouverte et collaborative est centré sur la mesure de pertes en commutation, son automatisation et sa précision, incluant les aspects hardware, de commande et d'instrumentation. A l'image des travaux passés tels que \cite{Opposition1,Opposition2,Opposition3,Opposition4}, la mesure des pertes en commutation par la mise en opposition de deux cellules de commutation permet une mesure précise mais indirecte : les 2 cellules doivent fonctionner en régime permanent, avec la mesure de la puissance moyenne fournie par l'alimentation DC. Ceci induit un fonctionnement à point nominal, générateur de pertes et alors d'auto-échauffement des semiconducteurs. Il est donc indispensable d'effectuer cette mesure le plus rapidement possible, d'où l'idée d'une automatisation. D'autre part, les ordres de commutation doivent être synchronisés, précis, et avec une résolution qui tend à se réduire, particulièrement dans le cas de transistors de puissance SiC et GaN. La résolution temporelle pour les temps morts et la régulation éventuelle du courant lors de la mise en opposition doit être particulièrement réduite, par exemple à l'échelle de la nanoseconde.

\begin{figure}[!hb]
	\begin{center}
		\includegraphics[width=0.8\columnwidth]{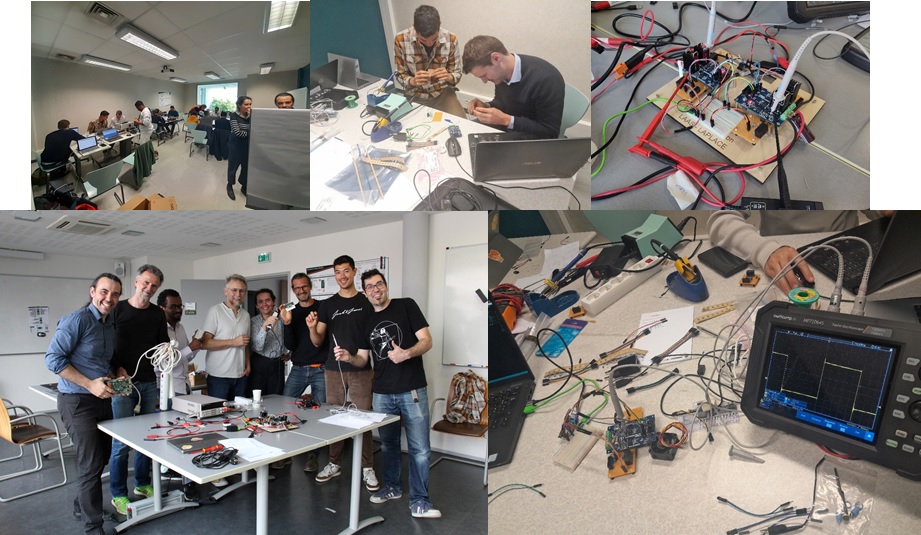}
	\end{center}
	\caption{Une aventure humaine pour la science collaborative et ouverte.}
	\label{fig_1}
\end{figure}

Ainsi, la mise en \oe uvre d'un tel banc de mesure automatisé des pertes en commutations est une tâche complexe, nécessitant de nombreuses compétences. Il peut être difficile de maîtriser tous les aspects nécessaires à l'instrumentation, la commande, la métrologie, les topologies de convertisseurs statiques, l'automatisation. Ce constat est particulièrement avéré pour les doctorant.e.s, mais aussi pour les ingénieurs et chercheurs de domaines connexes. Le travail collaboratif en science ouverte que nous avons mené nous a permis de rassembler un groupe de travail permettant de s'attaquer à ce problème. Ce travail a été initié par Nicolas Rouger et Luiz Villa qui, suite à des échanges préliminaires sur le sujet, ont préparé puis conduit 2 jours de hackathon au printemps 2024. Ces 2 jours ont regroupé plus de 20 participant.e.s (figure \ref{fig_1}), et ont été suivis par des séances mensuelles pendant 8 mois, avec celles et ceux qui souhaitaient aller plus loin et finir les travaux.

L'article est organisé comme suit : la section \ref{sec:enjeux} décrit plus en détails le fonctionnement du banc automatisé mis en oeuvre pour la mesure de pertes en commutation par opposition. La section \ref{sec:resultats} présente quelques résultats expérimentaux qui ont d'ores et déjà pu être obtenus. Les conclusions et perspectives de ce travail sont données en section \ref{sec:conclusion}.

\section{Enjeux techniques et scientifiques}
\label{sec:enjeux}
\begin{figure}[h]
	\begin{center}
		\includegraphics[width=0.9\columnwidth]{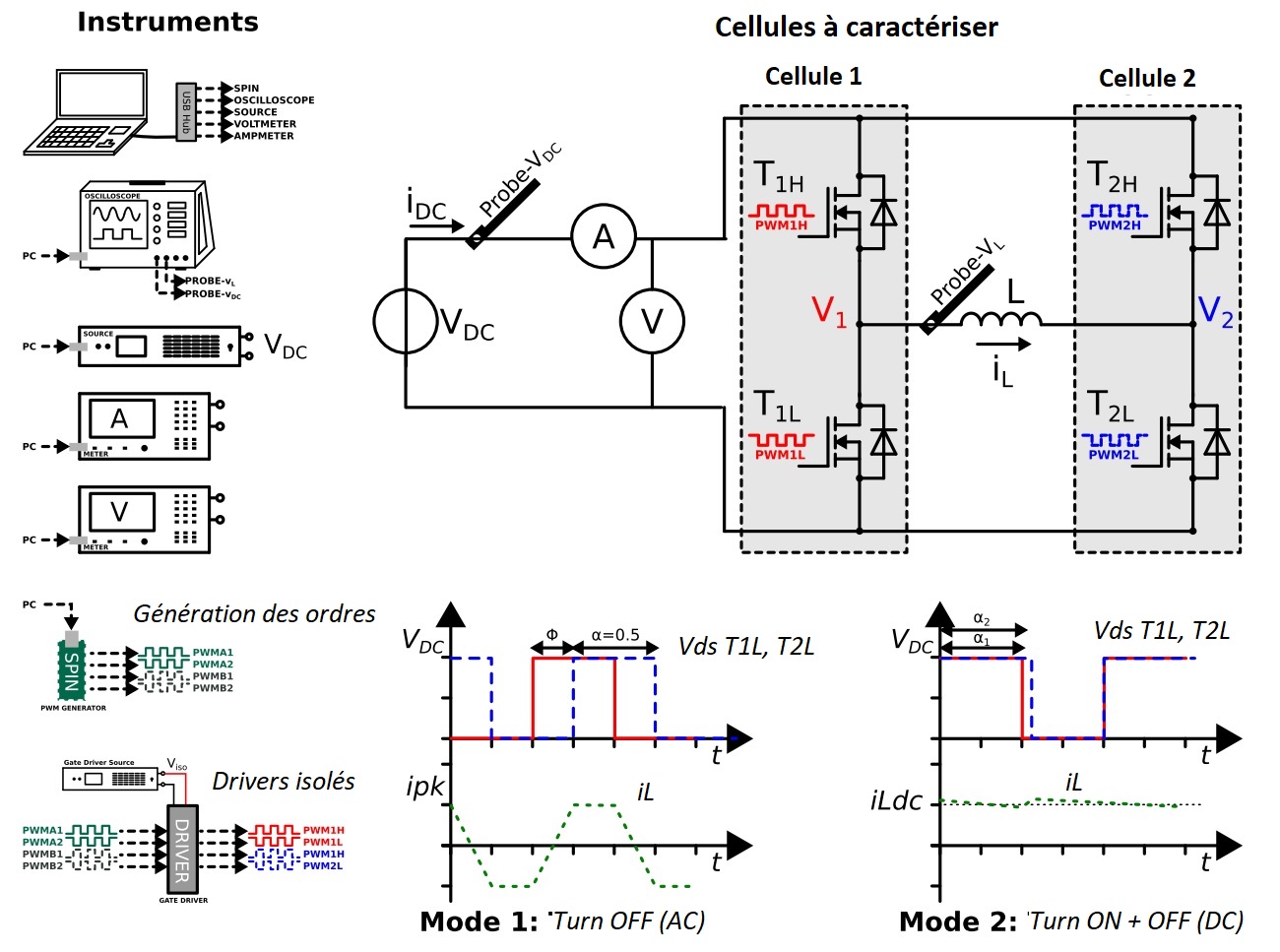}
	\end{center}
	\caption{Schéma de principe de la méthode d'opposition, instruments et organes principaux associés. Les deux modes de pilotage sont présentés.}
	\label{fig_2}
\end{figure}

\begin{figure}[h]
    \begin{center}
        \includegraphics[width=0.9\columnwidth]{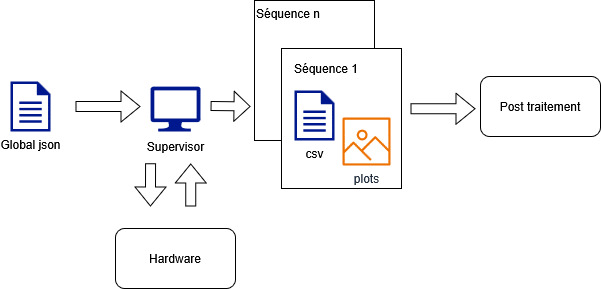}
    \end{center}
	\caption{Fonctionnement du banc automatisé de mesure de pertes en commutation par opposition.}
	\label{fig:fonctionnement}
\end{figure}

La figure \ref{fig_2} présente les 3 aspects considérés pour la mise en oeuvre du banc de  mise en opposition et la mesure automatisée des pertes en commutation: 
\begin{enumerate}
    \item Superviseur et instrumentation,
    \item Microcontrôleur et génération des ordres de commandes et des signaux de déclenchement,
    \item Hardware de puissance, incluant les commandes rapprochées et leurs alimentations.
\end{enumerate}
Ces différents volets sont détaillés dans les paragraphes suivants et fonctionnent ensemble comme indiqué sur la Figure \ref{fig:fonctionnement}. Un fichier de configuration définit les séquences de test. Chaque séquence correspond à un point de fonctionnement (défini par la tension $V_{DC}$), pour lequel un balayage sur le déphasage puis sur les rapports cycliques sera effectué. Un autre niveau de balayage a été fait sur la fréquence de commutation.

Lors du hackathon, nous avons défini un cahier des charges, avec les contraintes pré-identifées, et les participant.e.s se sont répartis sur les 3 aspects scientifiques et techniques. La contrainte que nous nous sommes fixés était d'utiliser et développer uniquement des outils en science ouverte, open source, à l'exception de l'utilisation de l'outil LTspice pour les simulations électriques.

\subsection{Superviseur et contrôle automatique des appareils}
Le superviseur en Python est en charge de la gestion complète et automatisée des essais. Il assure la configuration et pilotage des instruments en SCPI via PyVISA comme suit: 
\begin{itemize}
    \item \textbf{Source :} la source de tension DC est activée puis augmentée progressivement en rampe, stabilisée puis déchargée et coupée;
    \item \textbf{Mesures :} deux multimètres numériques servent à la mesure du courant moyen DC et de la tension moyenne DC (et par suite de la puissance moyenne), stockée dans leur mémoire pour réduire les temps de communication, et l'oscilloscope utilise sa mémoire segmentée afin d'acquérir rapidement chaque séquence d'essai.
\end{itemize}
Afin d'assurer la synchronisation des commandes et des mesures venant de multiples appareils, le superviseur ne fait que préparer la configuration de chaque instrument de façon séquentielle, sans déclencher les mesures. C'est un signal de synchronisation commun hardware qui déclenche l'acquisition des signaux à l'oscilloscope, les mesures, qui sont stockées de façon autonome dans les mémoires des appareils. A la fin de l'essai le contenu des mémoires est transférée de façon séquentielle et enregistré sous un format brut pour être traité. 

Un des enjeux clés du superviseur est d'assurer une grande précision des mesures moyennes, tout en réduisant le temps de fonctionnement du convertisseur, et donc la montée éventuelle en température des composants de puissance par auto-échauffement.

\textit{Remarque (Création du fichier de configuration):}
Pour aider à préparer, automatiser et traiter les essais, un script dédié à la génération de fichiers de configuration JSON\footnote{Le format JSON (\emph{JavaScript Object Notation}) est un format structuré utilisé pour stocker des données. Il contient des paires clé-valeurs avec pour avantage d'être lisible par l'homme et par la machine, facilitant le traitement informatique.} a été créé. Cela permet notamment la préparation d'un balayage sur plusieurs points de fonctionnement (valeur(s) de tension de bus DC, valeur(s) de courant max, valeur(s) de fréquence de commutation et autres paramètres tels que temps mort). Il est également possible de configurer un point de fonctionnement spécifique pour la validation des formes d'ondes avant l'essai automatique.  

\subsection{Génération automatique des signaux de commande}
Pour chaque point de fonctionnement considéré dans le fichier de configuration, la séquence de test consiste à utiliser deux modes de pilotage pour extraire séparément les pertes de blocage et d'amorçage : 
\begin{itemize}
    \item Ecart de déphasage entre les deux cellules pour un rapport cyclique fixe et identique.
    \item Ecart de rapport cyclique entre les deux cellules pour un déphasage nul.
\end{itemize}
Dans le travail effectué, chaque test d'un écart (de déphasage ou de rapport cyclique) a une durée fixe durant laquelle plusieurs mesures sont effectuées. L'implémentation est réalisée grâce au contrôleur de puissance OwnTech (SPIN), qui permet de générer les signaux synchronisés sur la base d'une horloge de référence pouvant aller jusqu'à 5.4 GHz. Au début du hackathon, il n'était pas encore possible de piloter la génération de signaux par une interface unique en Python et utilisant la communication série déjà en place.

\subsection{Aspects hardware}
\begin{figure}[!hb]
	\begin{center}
		\includegraphics[width=0.7\columnwidth]{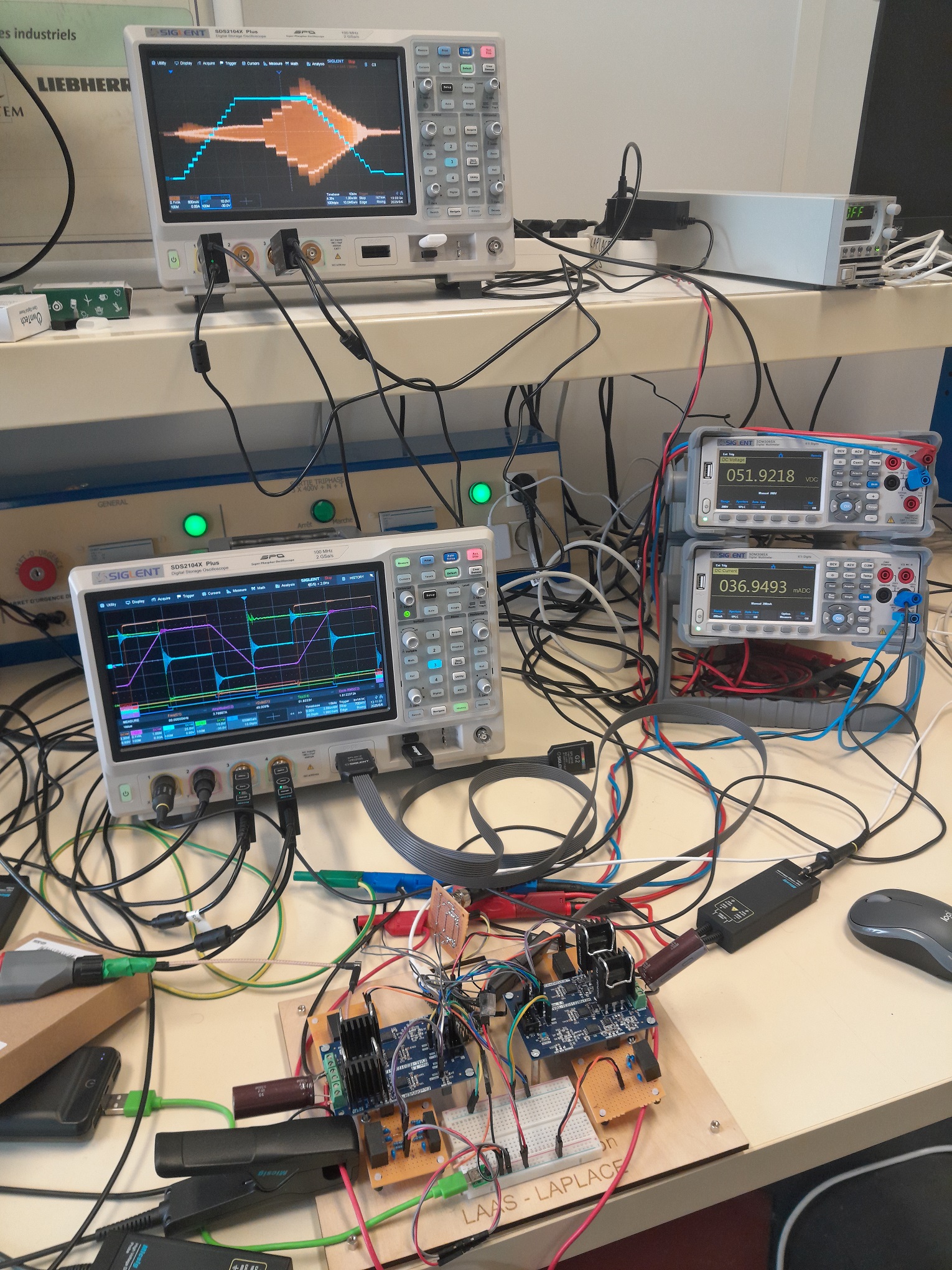}
	\end{center}
	\caption{Photographie du banc mis en oeuvre.}
	\label{fig_BENCH}
\end{figure}

La figure \ref{fig_BENCH} présente la mise en opposition de 2 cellules de commutation, initialement sans transistors de puissance ni alimentations isolées des quatre commandes rapprochées (cartes d'évaluation de demi-pont). 


Afin de pouvoir être réalisable dans le cadre d'un hackathon initial sur 48 heures, nous ne pouvions envisager la réalisation complète d'un hardware de puissance. Cependant, nous avions jugé important d'aborder le choix des composants, leurs commandes, et la soudure de composants dans le cadre de ce travail. Aussi, nous avons retenu en amont une carte d'évaluation de puissance, pouvant fonctionner à haute tension (600 V) et fort courant (> 20 A). La carte d'évaluation est EVAL-1ED3122Mx12H (Infineon), incluant une cellule de commutation "half bridge", avec les drivers isolés, mais sans les composants de puissance (T0-247-3) ni les alimentations isolées. Ces cartes d'évaluation sont plutôt destinées à des essais de double impulsion et non pas de commutation. L'objet du hackathon a alors été dans un premier temps de souder les composants manquants et mettre en place une simulation de l'architecture sur LTspice. Le choix a été fait d'utiliser des MOSFET Si (200 V, 20 A), afin d'avoir des commutations rapides même pour des tensions de bus DC réduites à 48 V. Cependant, il est possible de remplacer les composants, et plus globalement, le hardware n'est ici qu'un outil pour la mise en place complète de l'approche générique. Par la suite, le banc a été amélioré lors des séances mensuelles, pour donner la figure \ref{fig_BENCH}.
\section{Résultats expérimentaux}
\label{sec:resultats}
A ce stade, nous avons pu valider le fonctionnement du hardware, ainsi que la génération automatique des plans d'expérience et leur post traitement. Les sous parties vont présenter quelques exemples de résultats, à titre d'illustration de notre approche et des enjeux.

\subsection{Séquence d'essai}
\begin{figure}[!ht]
	\begin{center}
		\includegraphics[width=0.9\columnwidth]{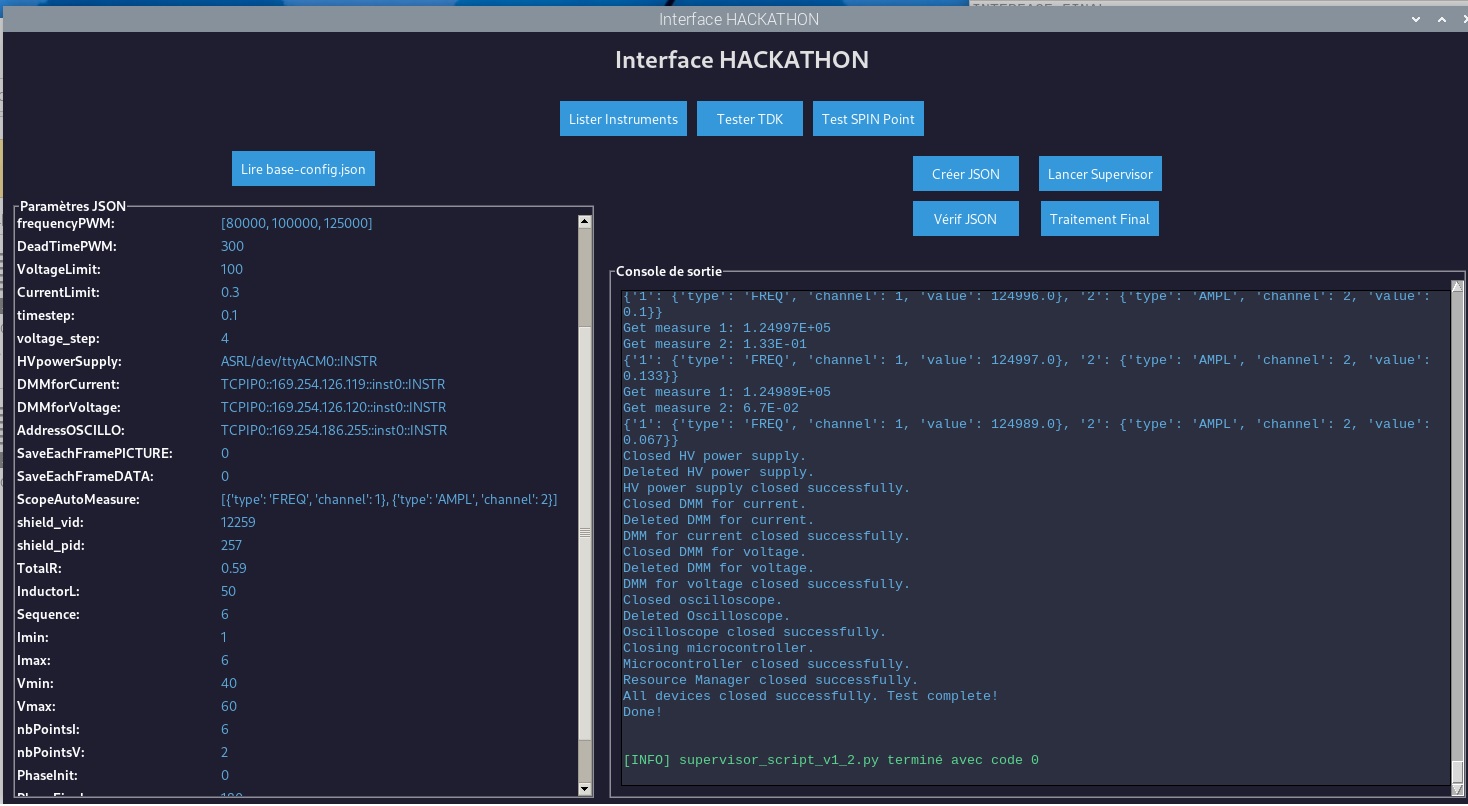}
	\end{center}
	\caption{Vue de l'interface graphique qui permet de réaliser toutes les opérations}
	\label{fig_Interface}
\end{figure}

Une interface graphique simple mais efficace a été réalisée en Python, permettant de lire le fichier json de configuration, qui inclue des paramètres communs (e.g. adresses IP des appareils, paramètres de communication) et des paramètres spécifiques aux plans d'expériences à réaliser. La figure \ref{fig_Interface} présente une capture de cette interface graphique. Il est aussi possible, en préalable aux plans d'expérience automatisés, de tester un point de fonctionnement unique, ainsi que le pilotage de la source de tension DC.

\begin{figure}[!ht]
	\begin{center}
		a)\includegraphics[width=0.8\columnwidth]{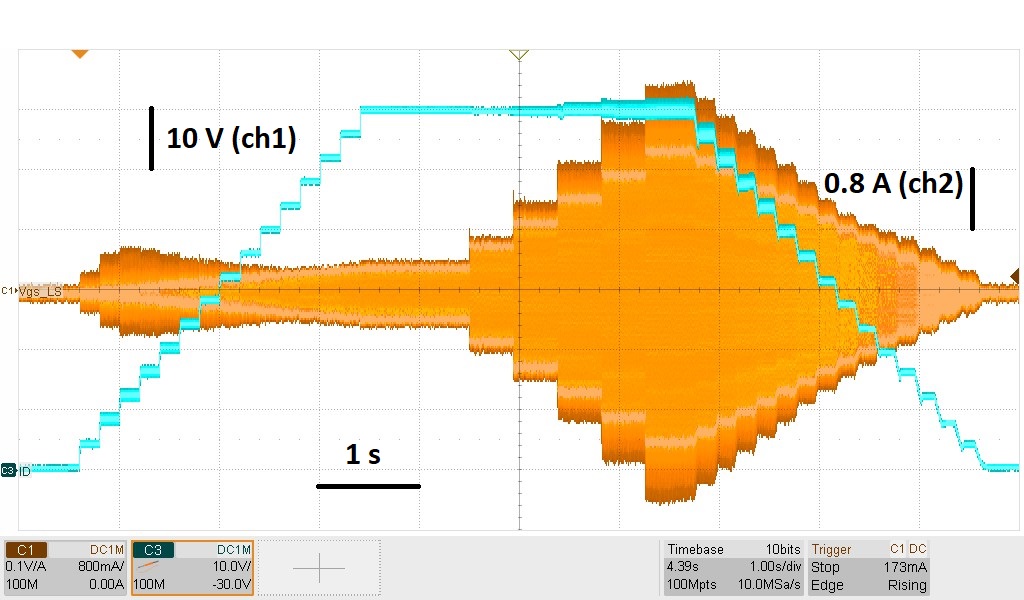}
		
		b)\includegraphics[width=0.8\columnwidth]{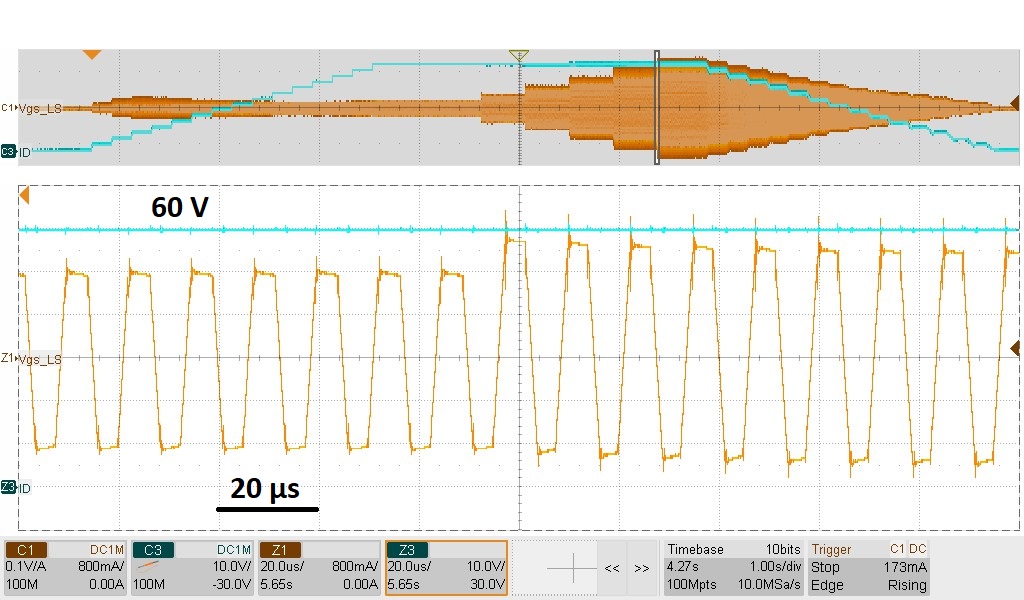}
	\end{center}
	\caption{Essai sur un balayage à fréquence fixe, avec un balayage du courant, précédé d'une montée progressive de la tension de bus. a) capture de l'essai complet, sur large plage temporelle. b) Zoom sur un changement automatique de point de fonctionnement en courant.}
	\label{fig_ZZZZZ}
\end{figure}

La figure \ref{fig_ZZZZZ} présente une capture des signaux clés, pour un  fonctionnement à 80 kHz, une tension de bus DC à 60 V et un courant crête proche de 3 A. Les transistors utilisés sont des MOSFETs Silicium 200 V. 

\begin{figure}[!ht]
	\begin{center}
		a)\includegraphics[width=0.9\columnwidth]{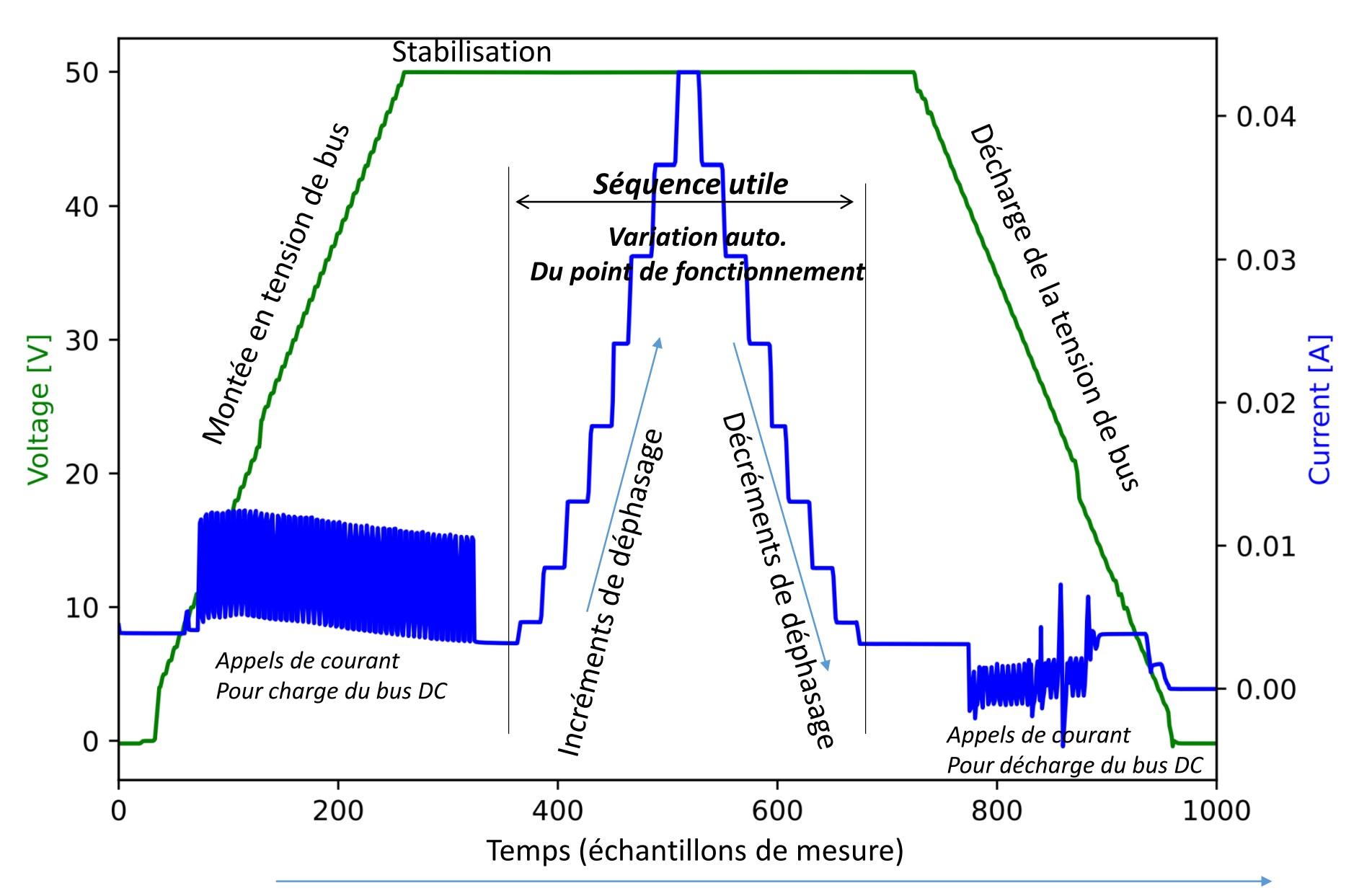}
		
		b)\includegraphics[width=0.9\columnwidth]{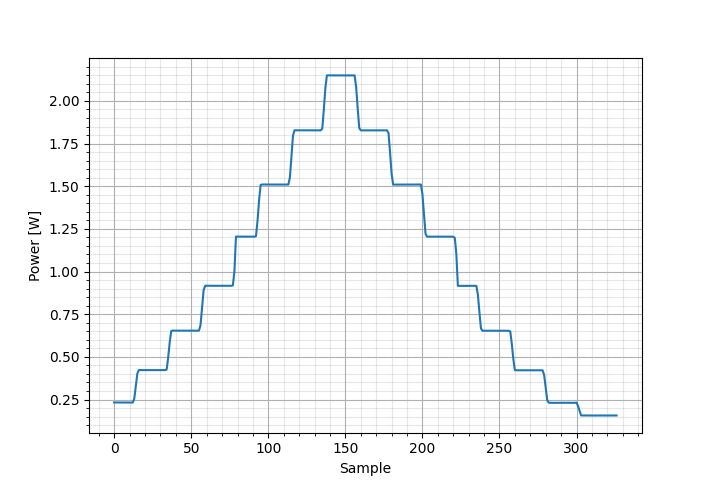}
	\end{center}
	\caption{Mesures acquises pendant un profil d'essai automatisé et préliminaire. a) Tension de bus (vert) et courant DC (bleu), b) Puissance moyenne dans la séquence utile de variation automatique de déphasage.}
	\label{fig_5}
\end{figure}

\begin{figure}[!ht]
	\begin{center}

		\includegraphics[width=0.8\columnwidth]{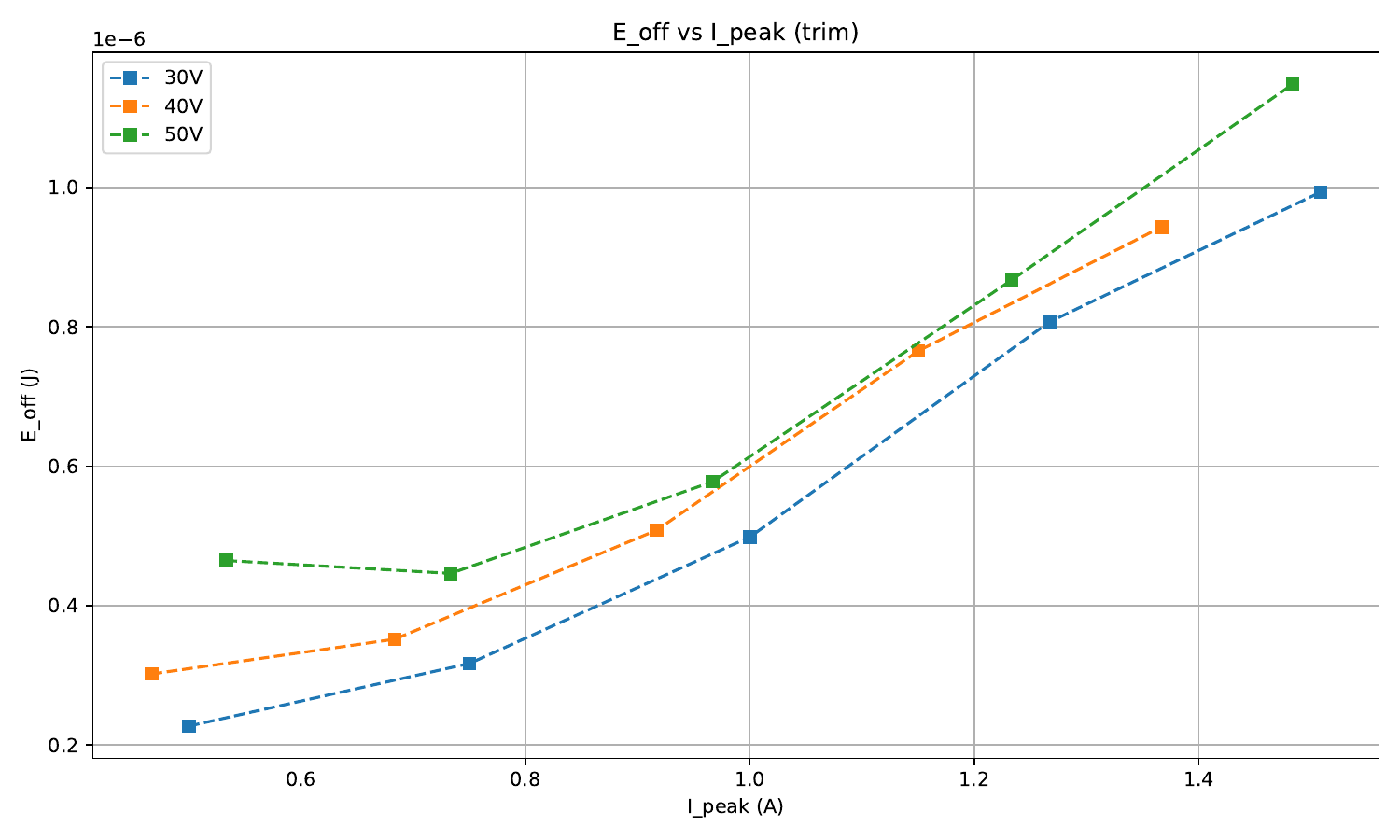}
	\end{center}
	\caption{Résultats issus de 2 plans d'expérience automatisés pour l'extraction automatique des énergies dissipées au blocage: balayage de 3 tensions de bus (30,40,50V) et 5 niveaux de courants commutés (de 0.5 A à 1.5 A).}
	\label{fig_ResultatPlanExp}
\end{figure}

Les pertes totales sont mesurées, et un soin particulier a été apporté à la synchronisation des mesures: la figure \ref{fig_5} présente un autre essai avec les résultats de mesure de courant et tension DC, pour un balayage de déphasage, et par suite de courant maximal commuté, ainsi que l'extraction des pertes totales. Ce profil d'essai peut ensuite être reproduit pour d'autres paramètres, tels que la tension de bus, la fréquence de commutation, la durée des temps morts. La cible est désormais de réaliser des mesures précises de tension et courant moyennés entre 1 ms et 20 ms, et faire le lien avec l'auto-échauffement associé lors de chacun des profils d'essais. La figure \ref{fig_ResultatPlanExp} présente l'extraction automatique des énergies de commutation au blocage.

\subsection{Zoom sur un point de fonctionnement}

\begin{figure}[!ht]
	\begin{center}
		\includegraphics[width=0.9\columnwidth]{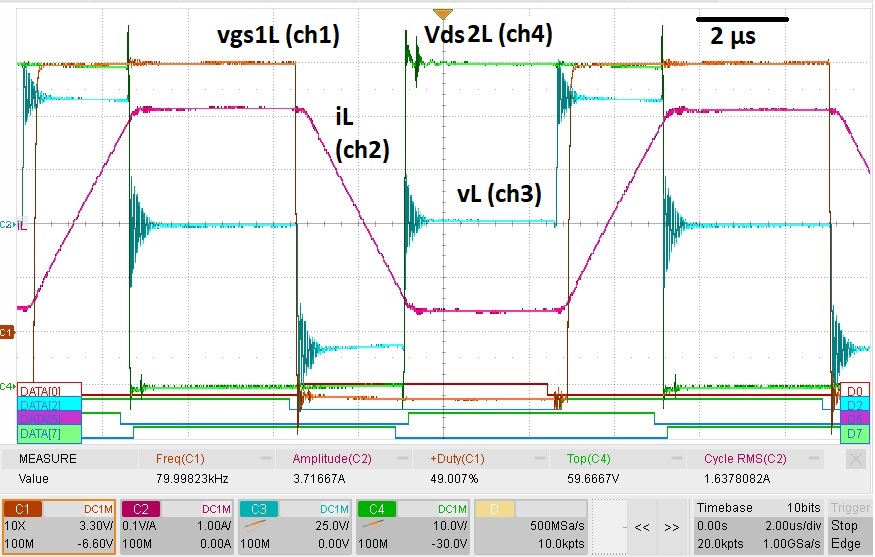}
	\end{center}
	\caption{Signaux principaux en fonctionnement, sur une période de découpage (mode déphasage). Paramètres: 60 V de tension de bus, fréquence découpage de 80 kHz, inductance de 50 $\mu$H, déphasage de 150$^{\circ}$, temps mort de 300 ns.}
	\label{fig_9}
\end{figure}

\begin{figure}[!ht]
	\begin{center}
		\includegraphics[width=0.9\columnwidth]{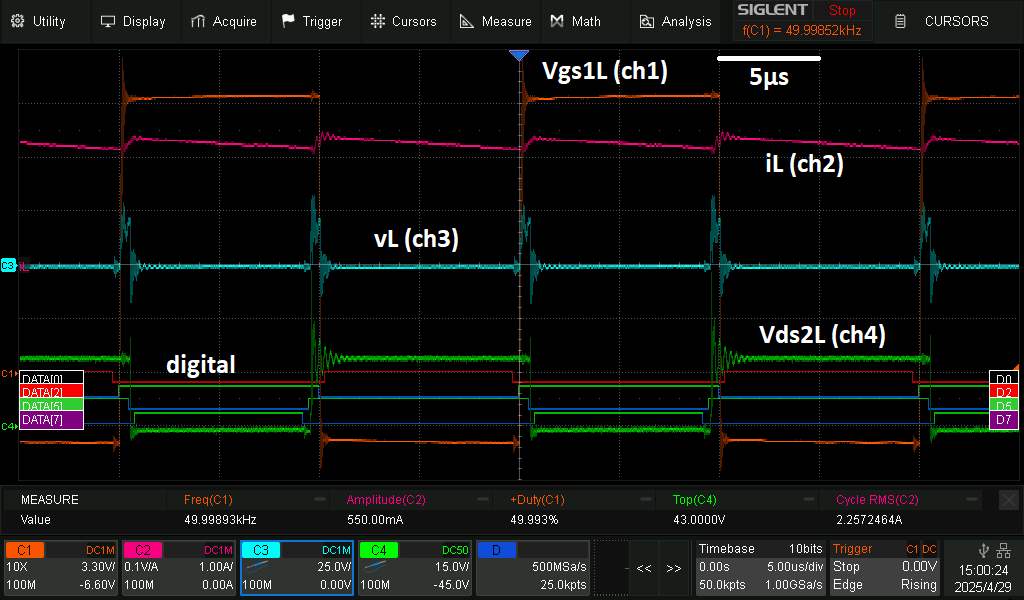}
	\end{center}
	\caption{Signaux principaux en fonctionnement, sur une période de découpage (mode écart de rapport cyclique). Paramètres: tension de bus réduite à 30 V, fréquence découpage de 50 kHz, inductance de 50 $\mu$H, temps mort de 300 ns, écart de rapport cyclique de 0.8 $\%$.}
	\label{fig_10}
\end{figure}

Des exemples de signaux acquis à l'échelle de la fréquence de commutation sont présentés sur les figures \ref{fig_9} et \ref{fig_10}. Ces signaux sont stockés dans la mémoire de l'oscilloscope, et les mesures sont automatiques et extraites après les essais, afin de comparer les grandeurs mesurées aux valeurs pré définies dans le plan d'expérience (e.g. courant maximal ou courant moyen). L'utilisateur peut récupérer toutes les images automatiquement, ainsi que les fichiers de points et les mesures de l'oscilloscope, en complément des grandeurs nécessaires acquises dans les deux multimètres numériques.

\section{Science ouverte et collaborative}
\label{sec:science_ouverte}
La forge Github à l'adresse \cite{Repo} regroupe les éléments utiles : le superviseur en python, ainsi que la configuration et pilotage du microcontrôleur, des instruments, et une partie du post-traitement des données et l'interface graphique (GPL v3). Les fonctions de génération des signaux de commande se sont appuyés sur les travaux déjà disponibles en science ouverte de la fondation OWNTECH, avec des développements supplémentaires sur la configuration par le superviseur Python des mode de fonctionnement et la génération du signal de déclenchement. 

\begin{figure}[!ht]
	\begin{center}
		\includegraphics[width=0.5\columnwidth]{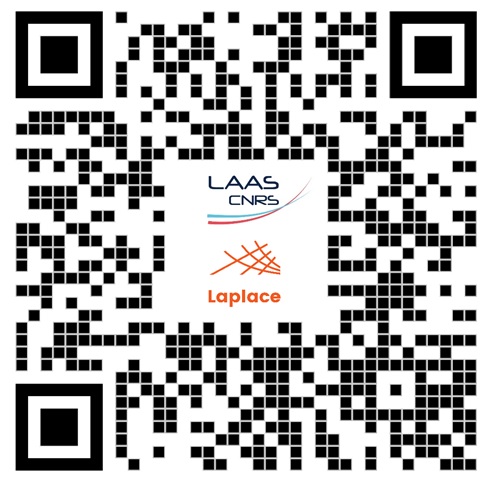}
	\end{center}
	\caption{Accès vers GITHUB. Actuellement, les branches "release candidate" et "clean up" sont les plus avancées, avant une release bientôt disponible sur la branche "master". }
	\label{fig_ZZZrezrezrzerzeZZ}
\end{figure}
\section{Conclusions et perspectives}
\label{sec:conclusion}

Nous avons présenté dans les grandes lignes notre démarche de science ouverte et collaborative, afin de mettre en place un banc automatisé de mesure de pertes en commutation par opposition. Le banc est fonctionnel, ses codes sont ouverts et disponibles, et nous avons déjà commencé à communiquer sur ces travaux. \\

\begin{figure}[h]
	\begin{center}
	    \includegraphics[width=\columnwidth]{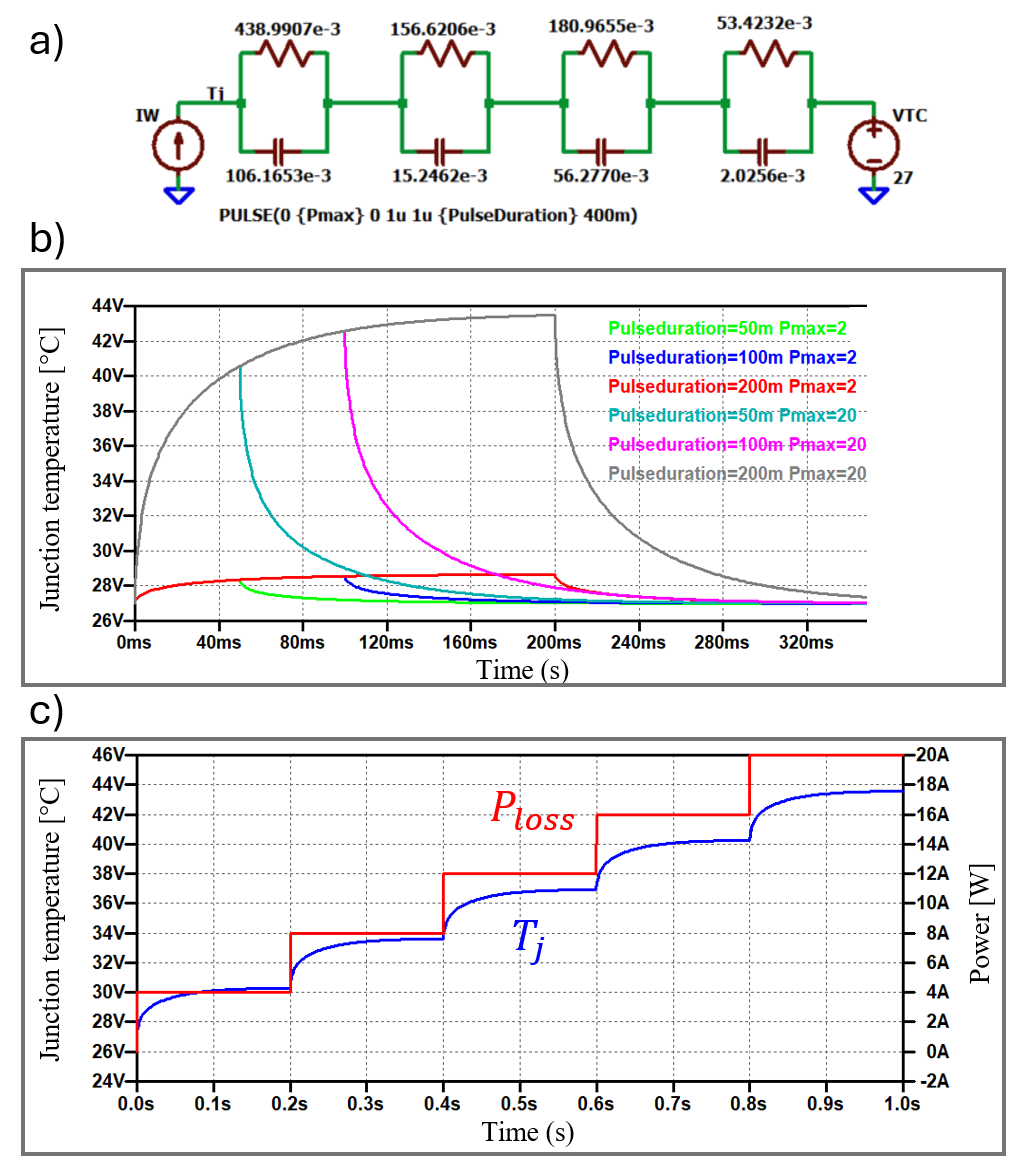}
	\end{center}
	\caption{Simulation électro-thermique en réponse à des échelons de pertes. a) schéma, b) impact de la durée du palier de pertes sur la température de jonction (2 ou 20 Watts, durées de 50,100 ou 200 ms). c) Evolution de la température de jonction lors de plusieurs séquences cumulées, de 0 à 20 W par pas de 4W et de durée de 200 ms.}
	\label{fig_Thermique}
\end{figure}

Les perspectives de ce travail porteront sur l'obtention de données (Si, SiC, GaN) et leur mise à disposition en accès ouvert via le même répertoire \cite{Repo}. Une étape supplémentaire de calibration automatique sera également proposé pour déterminer les paramètres du circuit (ex: L et sa Rdc, RdsON MOS, Rac inductance). Cette étape de pré calibration nécessite par exemple l'utilisation d'une commande sans découpage, avec limitation du courant, afin d'extraire les résistance totale du système, ce qui est utile pour la bonne estimation des pertes en conduction. Le banc automatisé présenté ici sera également amélioré afin de repousser ses limitations, en particulier sur les aspects thermiques. La figure \ref{fig_Thermique} montre 
les résultats des simulations électro-thermiques effectuées sur LTspice : l'élévation de la température de jonction à des échelons de pertes montre que les mesures effectuées dans le cadre de ce travail sont affectées. Il serait possible d'optimiser les séquences de test pour améliorer ce point, et/ou de prendre en compte l'auto-échauffement en compte lors du post-traitement.\\

La précision de mesure doit être aussi analysée, et comparée avec d'autres méthodes. Le compromis notamment sur la précision de la puissance moyenne et la réduction de la durée des paliers de point de fonctionnement doit être finement étudié. \\

Du point de vue des objectifs initiaux du hackathon, nous pensons les avoir atteint: un superviseur unique en Python est capable de piloter les instruments, générer les ordres de commande, récupérer les données et initier le post traitement. Un point important pour la suite: prévoir des mises en sécurité, notamment en cas de problème durant les plans d'expériences, tels que la perte de la communication avec l'alimentation, ainsi que l'automatisation du paramétrage des calibres (multimètres, oscilloscopes), selon les données pré définies dans le plan d'expérience.


\section*{Remerciements}
Nous remercions tout\textperiodcentered e\textperiodcentered s les participant\textperiodcentered e\textperiodcentered s au hackathon et sa suite mensuelle, "open power tuesday", ainsi que les personnes ayant contribué aux niveaux techniques et logistiques, au Laplace, LAAS et au sein d'Owntech. \emph{Note sur l'utilisation de l'IA générative}: nous avons utilisé de façon ponctuelle l'aide d'outils tels que Chat GPT et Mistral AI, afin d'avoir certaines routines de base, ou pistes de correction, ainsi que sur certaines briques de développement n'ayant pas de lien direct avec la méthodologie présentée ici.

\end{document}